\documentclass[pra,
a4paper,
showpacs,
twocolumn,
superscriptaddress]{revtex4}
\usepackage{amssymb}
\usepackage{amsmath}
\usepackage{amsfonts}
\usepackage{braket}
\usepackage{graphicx}
\usepackage{bm}
\usepackage{color}
\usepackage{multirow}
\usepackage{natbib}
\usepackage{hyperref}
\usepackage[normalem]{ulem}
\usepackage{verbatim}
\def \be {\begin{equation}}
\def \ee {\end{equation}}
\def \bea {\begin{align}}
\def \eea {\end{align}}
\def \p {\partial}
\def \BEA {\begin{eqnarray}}
\def \EEA {\end{eqnarray}}

\def \BC {\begin{cases}}
\def \EC {\end{cases}}

\def \be {\begin{equation}}
\def \ee {\end{equation}}
\def \bea {\begin{align}}
\def \eea {\end{align}}
\def \p {\partial}
\def \BEA {\begin{eqnarray}}
\def \EEA {\end{eqnarray}}

\def \BC {\begin{cases}}
\def \EC {\end{cases}}
\usepackage{graphicx}
\usepackage{dcolumn}
\usepackage{bm}
\usepackage{amssymb}
\usepackage{amsmath}
\usepackage{verbatim}
\usepackage{color}
\usepackage{ulem}
\def \be {\begin{equation}}
\def \ee {\end{equation}}
\def \bea {\begin{align}}
\def \eea {\end{align}}
\def \p {\partial}
\def \BEA {\begin{eqnarray}}
\def \EEA {\end{eqnarray}}

\def \BC {\begin{cases}}
\def \EC {\end{cases}}

\begin{document}
\title
{
  Tunneling Aharonov-Bohm interferometer on helical edge states
}

\author{R.\,A.~Niyazov}
\affiliation{NRC ``Kurchatov Institute", Petersburg Nuclear Physics Institute, Gatchina
188300, Russia}
\affiliation{L. D.~Landau Institute for Theoretical Physics,142432,  Moscow Region, Chernogolovka,  Russia}

\author{D.\,N.~Aristov}
\affiliation{NRC ``Kurchatov Institute", Petersburg Nuclear Physics Institute, Gatchina
188300, Russia}
\affiliation{Institut f\"ur Nanotechnologie,  Karlsruhe Institute of Technology,
76021 Karlsruhe, Germany}
\affiliation{St.Petersburg State University, 7/9 Universitetskaya nab., 199034
St.~Petersburg, Russia}

\author{V.\,Yu.~Kachorovskii}
\affiliation{A. F.~Ioffe Physico-Technical Institute,
194021 St.~Petersburg, Russia}
\affiliation{L. D.~Landau Institute for Theoretical Physics,142432,  Moscow Region, Chernogolovka,  Russia}
\affiliation{Institut f\"ur Nanotechnologie,  Karlsruhe Institute of Technology,
76021 Karlsruhe, Germany}

\date{\today}

\date{\today}

\begin{abstract}
We discuss  transport through
an interferometer  formed by helical edge states tunnel-coupled to metallic  leads. We  focus on the experimentally relevant case of  relatively high temperature as compared to the level spacing  and discuss a response  of  the setup to  the  external   magnetic flux $\phi$  (measured in units of flux quantum) piercing the area encompassed by  the edge states. We  demonstrate that  tunneling  conductance of the interferometer is  structureless in   ballistic case but shows  sharp antiresonances,  as a function of  magnetic flux  $\phi$ --- with the period $1/2$--- in the presence of  magnetic impurity.    We interpret  the resonance behavior   as   a coherent enhancement of  backward scattering off magnetic impurity at integer and half-integer values  of flux, which is accompanied by suppression of the effective  scattering at other values of flux.  Both enhancement and suppression are  due to the interference of processes with multiple returns to magnetic impurity  after a number  of clockwise and counterclockwise revolutions around setup.  This phenomenon is similar to the well-known weak-localization-induced enhancement of backscattering in disordered systems.  The  quantum correction to the tunneling conductance is shown to be proportional  to flux-dependent  ``ballistic Cooperon''. The obtained results can be used for flux-tunable control of the magnetic disorder in  Aharonov-Bohm interferometers built on helical edge states.
\end{abstract}

\maketitle
\section{Introduction}

The  quantum interferometry is a rapidly growing area of fundamental research with a huge potential for applications in optics, electronics, and spintronics. 
One of the simplest realization of quantum electronic interferometer   is  a   ring-geometry setup
tunnel-coupled to   metallic leads.   Such a  device can be controlled by magnetic field due to the  Aharonov-Bohm (AB) effect \cite{bohm,aronov}.  The  AB interferometers  formed by  single or few ballistic quantum channels   are very attractive  both from a fundamental point of view as prime devices to probe quantum coherent phenomena and  in view of possible applications as  miniature and very sensitive sensors  of magnetic field.
Although
they
   have been studied in detail theoretically, their
practical implementation faces significant difficulties.
The complexity of creating of ballistic single- or few-channel interferometers based on conventional semiconductors, such as GaAs or Si, is connected with technological problems of manufacturing  one-dimensional clean systems.
The efficiency of quantum electronic  interferometers used in practice is limited by rather stringent requirements, for example, very low temperature for interferometers based on superconducting SQUIDs or the requirement of very strong magnetic fields for interferometers based on the edge states of the Quantum Hall Effect systems.

A promising opportunity for a technological breakthrough in this direction is associated with the discovery of
topological insulators, which are materials insulating in the bulk, but  exhibiting  conducting channels at the surface or at the boundaries. In particular, the two-dimensional topological insulator phase was predicted in HgTe quantum wells \cite{Kane2005,Kane2005a,predicted} and confirmed by direct   measurements of conductance of the  edge states \cite{confirmed} and  by  the experimental analysis of the non-local transport   \cite{Roth2009,Gusev2011,Brune2012,Kononov2015}.
These states  are
 one-dimensional helical channels where the electron spin projection is connected with its velocity, e.g. electrons traveling in one direction are characterized by spin ``up'', while electrons moving in the opposite direction are characterized by spin ``down''.
 Remarkably, the electron transport  via helical edge states is ideal, in the sense that electrons do not experience backscattering from conventional non-magnetic  impurities, similarly to what occurs in edge states of Quantum Hall Effect systems, but without invoking high magnetic fields (for detailed discussion  of properties of  helical edge states see Refs.~\cite{Hasan2010,Qi2011}).
 Hence, in the absence of magnetic disorder, the boundary states are  ballistic and the intereferometers constructed on such states are topologically protected from external perturbations. Due to this key advantage
the helical edge states   are very promising candidates for building blocks  in  quantum  spin-sensitive  interferometry.

   The topological insulators have become a hot topic in the last decade  (see Refs.\ \cite{Chu2009, Hsieh2009,Bardarson2010, Peng2010,Michetti2011, Buttiker2012,Masuda2012, Bardarson2013,
mirlin1, kvon2,Dutta2016,Lin2017,gornyi1} and references therein).
 Particularly, different manifestations of the  AB effect  in  topological insulators were discussed: the dependence of the longitudinal conductance of  nanoribbons  and nanowires  on the  magnetic  flux  piercing   their cross-section    was studied   \cite{Peng2010,Lin2017}; weak antilocalization was  investigated in the disordered topolgical insulators  and oscillations with magnetic flux with the period   equal to the half of the flux quantum  were predicted \cite{Bardarson2010,Bardarson2013}.   The AB effect was also discussed for    almost closed loops formed by curved edge states   \cite{Buttiker2012}.   Experimentally,  the AB oscillations were  observed in the magnetotransport  measurements of  transport (both local and nonlocal) in 2D topological insulators based on HgTe quantum wells \cite{KvonAB2015} and were explained  by coupling of helical edges to bulk puddles of charged carriers.

\begin{figure}
\includegraphics[width=0.7\columnwidth]{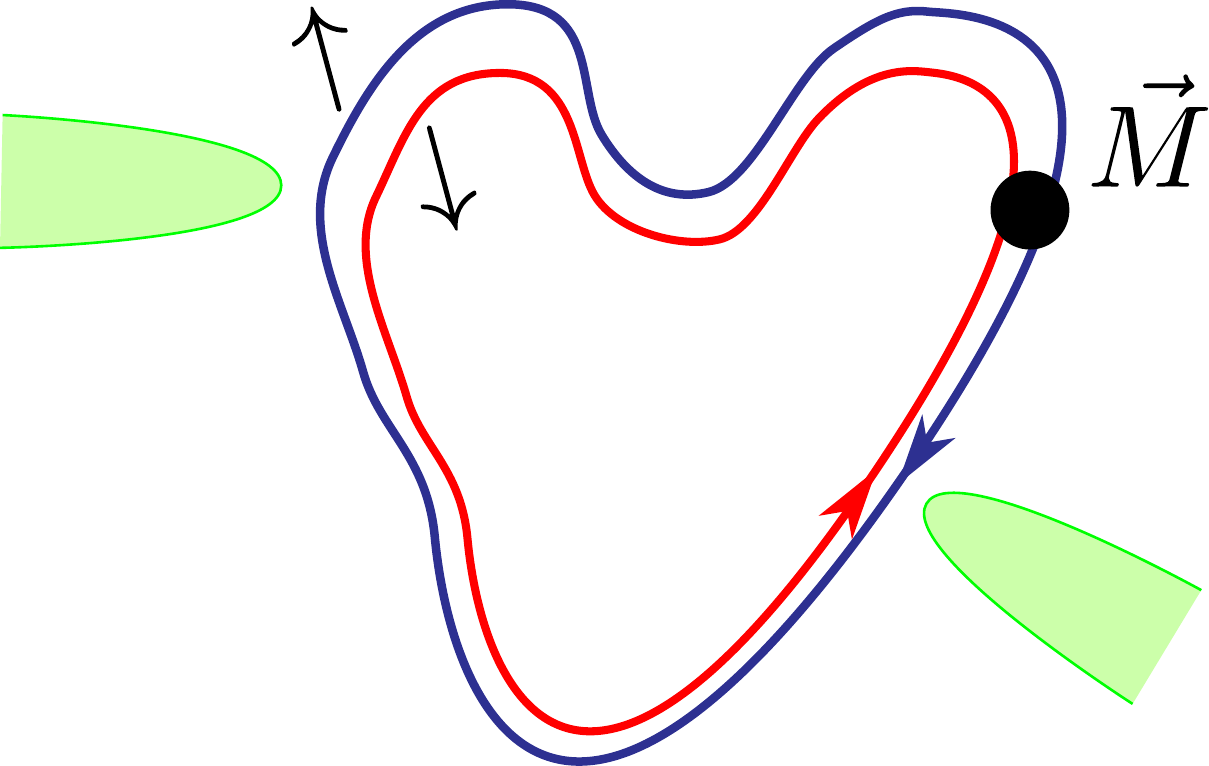}
\caption{\label{fig:densityring}
Interferometer formed by helical edge states tunnel-coupled to the metallic point contacts. The upper shoulder of the interferometer contains  magnetic impurity which serves as a backscattering center. }
\end{figure}

 The purpose of the current paper is to discuss a standard AB setup (see Fig.~\ref{fig:densityring}) focusing on the effect of the magnetic impurities.
We    consider an interferometer formed by  helical-edge states   of a 2D topological insulator   tunnel-coupled
 to leads. We assume that upper and lower shoulders of   interferometer have, respectively,   lengths $L_1$  and $ L_2.$
 We limit ourselves to discussion of  setup with  a single impurity placed into upper shoulder at the distance $s$ (along the edge) from the left contact. This case captures all essential physics of the problem and can be easily generalized for a  more general case  of arbitrary number of magnetic centers  provided that magnetic disorder is weak and the corresponding localization length  is large   compared to  interferometer circumference $L=L_1+L_2.$

Such AB interferometer built on the helical edge states was studied theoretically at zero temperature  \cite{Chu2009, Masuda2012,Dutta2016}.
 It is known, however, that the  form and shape  of the AB oscillations   strongly depend on the relation between temperature $T$  and level spacing  $\Delta$  connected with the finite size of the system (see
  a more detailed discussion below).
   In our case,  $\Delta=2\pi v_F/L$ is controlled by total length $L$ and  the Fermi velocity  $v_F.$
    For typical sample parameters, $L =10~ \mu$m,  and $v_F=10^7$ cm/s, we estimate  the level spacing $\Delta \approx 3 $ K.  Having in mind not too low   temperatures,
    we  focus on the experimentally relevant case,
\be T\gg \Delta  \label{TggD},\ee
and discuss dependence of the tunneling conductance $G$ of the setup on the     external   magnetic
flux $\Phi$ piercing the area encompassed by  edge states.
Simple estimates show that the flux quantum,  $\Phi_0=hc/e$, is achieved for rings of the above size in fields  $B\sim 3\mbox{ Oe}$,  well below the expected magnitude of the fields destroying the  edge states
\cite{Du2015,Zhang2014,Hu2016a}. Therefore, in what follows,  we neglect  the influence of the magnetic fields on the helical states.
 We will   demonstrate the existence of interference-induced effects, which are robust to the  temperature, i.e. survive under the condition \eqref{TggD},  and can therefore be obtained for relaxed experimental conditions.
Specifically, we  find that   $G$ is  structureless in   ballistic case but shows a sharp antiresonances,  as a function of $\phi=\Phi/\Phi_0$
 in the presence of
 magnetic impurities.
 Although
  similar antiresonances  are    known to arise in the  single-channel  rings made of conventional materials, 
  the  helical AB interferometer   shows   essentially  different  behavior  due to specific properties of the edge states.
  Most importantly,
  the effect is more  universal and robust to details of the setup, in particular, to the relation between $L_1$ and $L_2.$      
  Another difference  concerns the periodicity  of the function $G(\phi),$ which --- for the case of helical edge state interferometer --- obeys $$G(\phi+1/2)=G(\phi),$$ while  for interferometers made of conventional materials, this function is periodic with the period $1$.

  The  resonance behavior of conductance  arises
  due to  the interference contribution from  trajectories containing  multiple returns to the magnetic impurity after ballistic  revolutions around the setup  in clockwise and counterclockwise directions.  Such processes are  specific for closed geometry of the system  and  are absent in the infinite helical edge containing  backscattering  centers       (see discussion of  transport in latter system   in a number of recent publications \cite{Dolcini2011, Altshuler2013, Vayrynen2013,Vayrynen2014,Kimme2016,Vayrynen2016,kur3}).
   We notice, however, that bulk puddles of charged particles coupled to  infinite  helical edge \cite{Vayrynen2013,Vayrynen2014,Vayrynen2016}    (or curved edge states with  tunneling coupling between different points \cite{Dolcini2011,Buttiker2012})  can serve as mini-resonators and the edge transport can probe levels in these resonators. This  effect was used for interpretation of experimental results  on AB oscillations in non-local transport measurements in  HgTe  quantum well  \cite{KvonAB2015}.   What we claim here is that for almost closed AB interferometer with very weak tunneling coupling to the leads,  the circumference of the interferometer itself can serve as such a resonator and increase or decrease --- depending on the value of $\phi$ --- effective scattering  on the impurity.


Next, in order to compare our results with previously obtained ones, we  briefly summarize the  properties of the
AB interferometers formed by  conventional (non-helical)    {\it single- or few-channel  ballistic} channels, which were actively studied in last decades
\cite{butt,Gefen,Butt1,jagla,Yacoby5,kin,AB2,Chang,dmitriev,Shmakov2013,Dmitriev2015,SCS2017}.

For noninteracting ring, the tunneling conductance of the interferometer,
\be
G=N \frac{e^2}{h}     {\mathcal T}\, ,
\label{G}
\ee
is expressed in terms of energy-averaged tunneling transmission coefficient
$
    \mathcal T=-\int d\epsilon\, \mathcal T(\epsilon) \p_\epsilon f_F(\epsilon),
$
 where $f_F$ is the Fermi distribution function and $N$ is the number of the conducting  channels.
The tunneling conductance  is known to exhibit different types of oscillations.
Specifically,   $\mathcal T(\epsilon)$  oscillates   as a function of  energy of the tunneling electron $\epsilon$, having maxima at positions of quantum levels in the ring \cite{butt,Gefen,Butt1}. For low temperatures, $T\ll \Delta$, these oscillations  transform into oscillations of $G$ with the Fermi energy, which    are strongly affected by the  Coulomb blockade in the interacting case \cite{kin}.  In the opposite high temperature case, $T\gg \Delta,$  these oscillations are exponentially suppressed.

Magnetic field applied to the ring  leads to  AB oscillations ---periodic  oscillation  of   $G$    with the dimensionless magnetic flux $\phi$ piercing  the area encompassed by quantum channels. The period of the oscillations is given by the flux quantum $ \varDelta \phi=1.$
The shape and amplitude of  the  oscillations depend essentially on the strength of the tunneling coupling and on the relation between $T$ and $\Delta.$
  For  $T \ll \Delta$ and weak tunneling coupling  there are {\it two}  narrow {\it resonances}  in the dependence $G(\Phi)$ \cite{butt,Gefen, Butt1} within the interval $0<\phi<1.$
  The positions of the resonant  peaks depend on the electron Fermi energy \cite{butt} and on the strength of the electron-electron interaction \cite{kin}.  Remarkably,  the interference effects are not entirely suppressed with increasing the temperature, and the resonant behavior of $G(\Phi)$ survives
for the case $T\gg \Delta.$  However, this dependence changes qualitatively.
In particular, the high-temperature conductance of the  noninteracting  ballistic  ring with $L_1=L_2$   weakly   coupled to the contacts exhibits {\it single} sharp {\it antiresonance}   in the interval $0<\phi<1$ at $\phi = 1/2  $  \cite{jagla,dmitriev}.
The antiresonances at $\phi= 1/2+n$ ($n$ is integer number) are  broadened  by weak disorder \cite{Shmakov2013}.  The electron-electron interaction leads to  appearance of a fine structure  of the antiresonances: each antiresonance splits     into a series of  narrow peaks, whose widths
are governed  by dephasing \cite{dmitriev,Dmitriev2015}. Spin-charge separation in the interacting  spinful interferometer leads to the additional splitting of antiresonances because of existence of two types of excitations (charge and spin excitations) proparating with different velocities \cite{jagla,SCS2017}.

Additional physics comes into play in the presence of the spin-orbit (SO) interaction, which results in a   phase   acquired  by the spin part of the
electron wave
propagating  around the ring.
 This phase is added to AB phase and  leads to  the third type of the conductance  oscillations:  periodic oscillations of $G$ with the strength of the SO coupling, so  called  Aharonov-Casher (AC)  effect \cite{AC,Stone1,Stone}.
 The effect of SO on the performance of the AB interferometer   was intensively discussed  \cite{AC,Stone1,Stone,history1,history3,history4, citro2,citro1,kov, moldov,exp1,
 Nagasawa2012,Nagasawa2013} with the focus on   zero-temperature case.
The finite temperature effects were also analyzed both  numerically  \cite{history4,citro1,moldov}, and  analytically  \cite{Shmakov2012}.   In particular, it was found in Ref.~ \cite{Shmakov2012} that    AC effect is also robust to temperature:  for $T\gg \Delta,$ the antiresonances in the conductance are split  by SO coupling with the splitting distance proportional to the AC phase.

Since topological insulators are materials where SO interaction is strong, one might expect that the  flux dependence of $G$ in interferometers based on helical edge states would be very similar to the case of
single- or few-channel interferometers based on AC effect.
We show below that this is not the case. The antiresonances arising in the usual AB interferometers at high temperatures, $T \gg \Delta,$ both in absence and in the presence of SO,    turn out to be  very sensitive to the geometry of the problem
\cite{dmitriev, Shmakov2012, Shmakov2013, Dmitriev2015, SCS2017}.
In particular, their shape and width are strongly modified for interferometer with non-equal shoulders, $L_1\neq L_2$  \cite{app-dmitriev}.  By contrast, as we will see,   the antiresonances  in the  interferometers formed by helical edge states are  not  sensitive to geometry of the device: their  amplitude and width  do  not depend  on the relative lengths of the interferometer shoulders  and on  the position of the magnetic impurity.  Also, they do not change in the non-planar geometry of the sample. In the next sections, we present   both rigorous calculations and  a simple physical interpretation of these properties.
\section{Basic equations}
\label{basic}
We assume that   the helical edge states of  two-dimensional topological insulator are      tunnel-coupled  to   metallic nonmagnetic  point contacts.     We model metallic  contacts  by single-channel spinful wires, so that   electrons are injected into  the helical states through  so-called tunnel Y junctions.  We assume that the electrons in the metallic leads are not polarized.
%
 In this paper, we focus on the calculation of the tunneling conductance of the interferometer, whereas interesting effects related to spin-selective properties of the system  such as resonance rotation of initial spin polarization will be discussed elsewhere \cite{niyazov-to-be-published}.

\begin{figure}
\includegraphics[width=0.6\columnwidth]{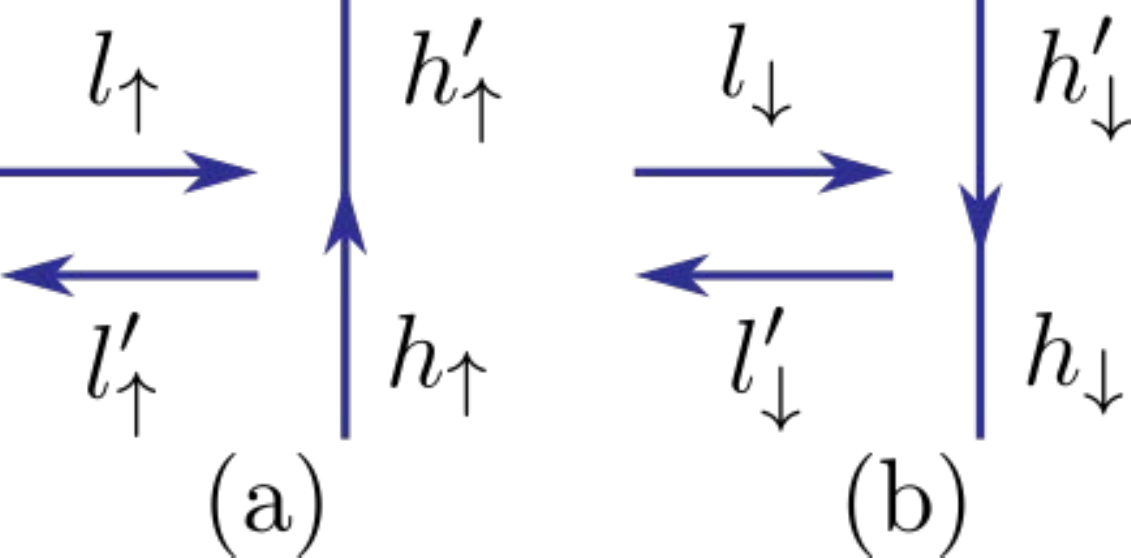}
\caption{\label{fig:pointcontact}
Point contact between the helical ring and the spinful wire.}
\end{figure}

Since we consider  nonmagnetic  leads,
 the different spin projections do not mix at the contacts. Thus, the point contact is described by the time-reversal $S$-matrix:
\begin{equation}
S=  \begin{pmatrix}
-t& r & 0 & 0 \\
r &  t & 0&0\\
0 & 0  &-t &r \\
 0 & 0 &r & t \\
\end{pmatrix}.
\end{equation}
This $S$-matrix connects incoming states $(l_\uparrow,h_\uparrow,l_\downarrow,h_\downarrow)$ with outgoing states $(l'_\uparrow,h'_\uparrow,l'_\downarrow,h'_\downarrow)$ (see Fig.~\ref{fig:pointcontact}). ``l'' stands for the leads, ``h'' stands for the helical edge. This notation corresponds to one in Ref.\ \cite{Aristov2016} after obvious  rearrangement of channels. The difference between the Fig.~\ref{fig:pointcontact}a  and Fig.~\ref{fig:pointcontact}b is the opposite propagation of electrons with different spins in the helical edge.

The  tunneling contact is characterized by two amplitudes $r$ and $t,$  obeying $|t|^2+|r|^2=1.$ Without a loss of generality, we assume that $t$  and  $r$ are real   and    express them through the  parameter  $\gamma,$
\begin{equation}
 r=\frac{2 \sqrt{\gamma}}{1+\gamma}, \quad  t=\frac{1-\gamma}{1+\gamma},
\label{rt}
\end{equation}
which  has a physical meaning of  tunneling transparency of the contact \cite {dmitriev,Dmitriev2015,Aristov2010}.


Let us calculate the phase acquired by  the electron wave  after  a full revolution around the setup shown in Fig.~\ref{fig:densityring}.   In order to make the physical picture more transparent,   we  consider first a  general case with two chiralities (direction of propagation) and two spins not necessarily aligned  with momentum (this case corresponds to a conventional single-channel spinful wire).
The  acquired phase
includes three terms:  a dynamical contribution $kL$  (here $k$  is  the electron wave vector),
magnetic phase $\pm 2\pi \phi$ and the Berry's phase $\pm \delta$~\cite{berry}, given by one half of the solid angle, subtended by spin direction during circumference of the interferometer.     The dynamical contribution depends on $L$ only and does not change its sign when changing the chirality  and spin projection.   By contrast,  the sign of magnetic phase is insensitive to spin but changes sign with changing the chirality. The Berry's phase changes sign both with changing the chirality and with changing the  spin (see Tab.~\ref{tab:phases}).
 For helical edge only two electron states are present, which are marked by boldface in the Table~\ref{tab:phases}.

 Analyzing corresponding phases we arrive at a   conclusion, which is of key importance for our  analysis.
 Information about  the geometrical structure of the edge states, in particular, about  curvature of the   edge and/or non-planar geometry, is encoded in the Berry's phase, but as we see, it
    is  simply   added to the dynamical phase, which implies  that
    amplitude  of  any process depends on $kL+\delta.$ This, in turn, means that tunneling conductance for a given energy (i.e. before thermal averaging) depends on the Berry's phase and is, therefore, sensitive to geometry of  the setup. However,  for $T\gg \Delta,$  the  thermal averaging  implies integration over $k $  within a   wide  interval, $ \delta k \sim T/\hbar v_F \gg 1/L,$  around the Fermi wave vector  $k_F.$ After changing integration variable,   $k+\delta/L \to k^\prime$,  the Berry's phase drops out with  the exponential precision.    This should be contrasted to the case of conventional interferometers with weak SO coupling, where the Berry's phase contributes to the Aharonov-Casher phase and  strongly effects both  $\mathcal T(\epsilon)$  and     energy-averaged transmission coefficient, $\mathcal T.$
     As a result, the conductance of the ring depends on the Berry phase as was first demonstrated in  Ref.~\cite{history1} for low temperature case, $T\ll\Delta,$ and then generalized for the high temperature regime, $T\gg \Delta,$
        in Ref.~\cite{Shmakov2012}.
     Physically,    dependence  of  $\mathcal T$  on $\delta$  in the case of weak SO coupling  and  $T\gg\Delta$ arises because
     the   electron wave with a given spin polarization can propagate  both clockwise and counterclockwise and  the  phase shift between such waves   with equal winding numbers, $n_1=n_2=n$, is given by $2(\phi+\delta) n$.
      Such processes are  absent in the  interferometer based on helical edge states, and, as a consequence,  in the latter case  $\mathcal T$ is  independent of $\delta$ in the high temperature regime.

 \begin{table}
\begin{tabular}{cccc}
& & \multicolumn{2}{|c|}{chirality}   \\
\cline{3-4}
& & \multicolumn{1}{|c|}{+} &\multicolumn{1}{c|}{--} \\
\hline
\multirow{2}{*}{spin}  & \multicolumn{1}{|c|}{$\uparrow$} & \multicolumn{1}{|c|}{ \boldmath$k L + \phi + \delta$}  & \multicolumn{1}{c|}{$k L - \phi - \delta$ }  \\ \cline{2-4}
 & \multicolumn{1}{|c|}{$\downarrow$} & \multicolumn{1}{|c|}{$k L + \phi - \delta$}  & \multicolumn{1}{c|}{ \boldmath$k L - \phi + \delta$} \\ \hline
\end{tabular}
\caption{\label{tab:phases} Phases of electron wave function after a full revolution in the arbitrary setup shown in Fig.\  \ref{fig:densityring}. }
\end{table}

 \section{ Tunneling conductance
 }

The tunneling  conductance of the setup under discussion
 is given by Eq.~\eqref{G} with $N=2$. We have  $
G=(2e^2/h)  \mathcal T,
 $
 where, for the case of  spin-unpolarized contacts, the
transmission coefficient can be represented as
  an average over incoming spin polarizations
 \be
 \mathcal T=\frac {\mathcal T_\uparrow + \mathcal T_\downarrow}2  \,.
\label{T}
 \ee

For     fully ballistic interferometer,    $\mathcal T$  does not contain interference term and hence is flux-independent.   Indeed, the right- and left- moving electrons  have opposite spin projections at any point, in particular at the point contact where they exit  the ring. Consequently, they  can not interfere in the tunneling process.    Next, we take into account that  condition \eqref{TggD}  provides that
 $\delta k\, L \gg 1.$  Due to this inequality, interference  contribution  coming from any  two trajectories of the same chirality and with different winding numbers $n$ and $m$ becomes exponentially small, $\propto \exp (-\delta k L|n-m|) ,$ after the thermal averaging. Neglecting such exponentially small terms, we arrive at the conclusion that only trajectories  with the same chirality and $n=m$ contribute.  In other words,  the  transmission coefficient of fully ballistic system  is given by purely classical contributions, while all the interference terms are suppressed.   Classical contribution from the trajectory with certain chirality and winding number $n$  to the transmission coefficient  is given by $\left |r\cdot t^{2n}\cdot r \right |^2,$  The summation over $n$ yields  $\mathcal T_\uparrow = \mathcal T_\downarrow = r^4/(1-t^4)$, and the use of \eqref{rt} leads to
 \be
 \mathcal T=\frac{2\gamma}{ 1+\gamma^2}.
 \ee
Here the case $\gamma \ll1 $ corresponds to weak tunnel coupling mostly considered in this paper, and fully open setup is described by $\gamma=1$.



\begin{figure}[ht]
\centerline{\includegraphics[width=0.8\linewidth]{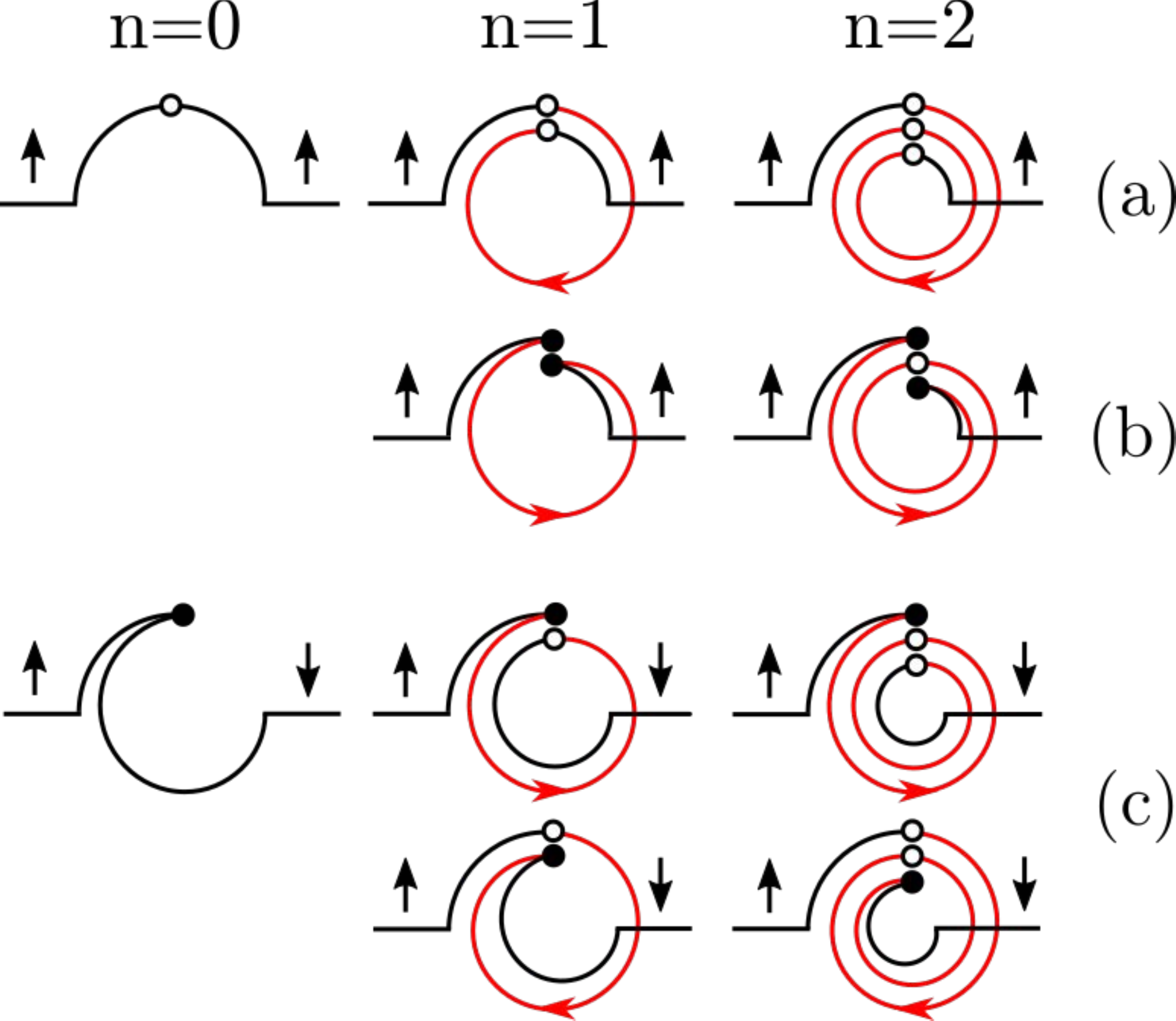}}
\caption{Processes with incoming spin up projection,  involving ballistic propagator as well as single and double backscattering processes at the magnetic impurity.  Backscattering events with the corresponding amplitude $i \sin \theta$  are marked by black dots. Forward scattering events by the magnetic impurity are shown by open circles and has amplitude $\cos\theta$.     The processes (a) and (b) with coinciding initial and final polarization can interfere.  The process (c) corresponds to spin-flip tunneling process.   A number of returns to the magnetic impurity denoted by $n$. The processes shown in (a) and (b) with equal $n$  correspond to $n$ revolutions around the ring in   opposite directions.      }
\label{proc1}
\end{figure}

\begin{figure}[ht]
\centerline{\includegraphics[width=\linewidth]{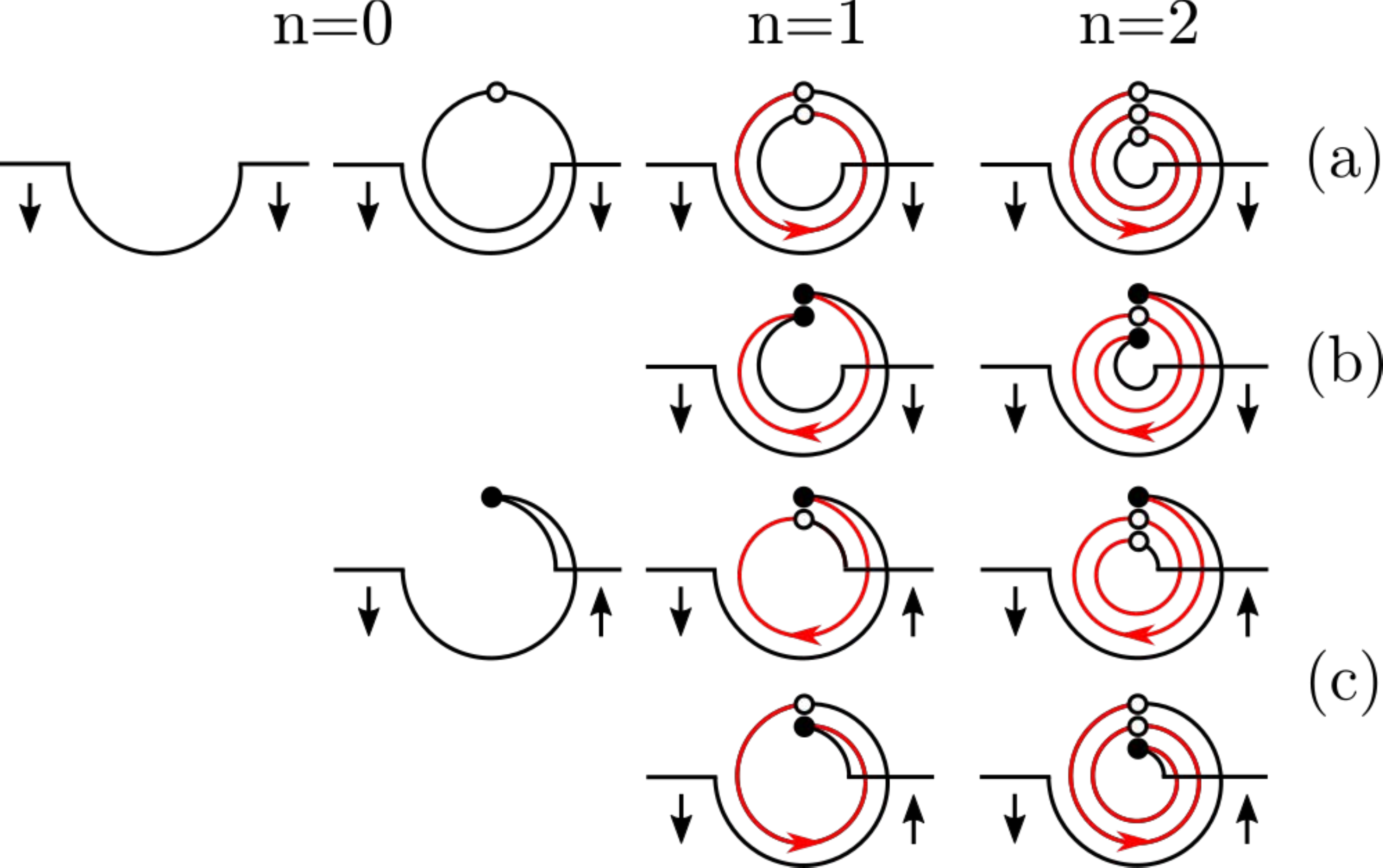}}
\caption{The same as in Fig.~\ref{proc1} but with spin down projection.}
\label{proc2}
\end{figure}

 Let us now demonstrate that the magnetic impurity induces quantum interference corrections to the conductance.
We will discuss the case of classical impurity with large magnetic moment
 $\mathbf M$ ($M\gg 1$).
Then, in the first approximation, one can  neglect  feedback  effect related to the dynamics of $\mathbf M$ caused by  exchange  interaction with the ensemble of  right- and left-moving  electrons (for infinite helical edge  this  effect was discussed in Ref.~\cite{kur3}).

 In this section, we assume that impurity is static and
 is described by the scattering matrix of general form
\be
\hat S_M=\begin{pmatrix}
 \cos \theta &  i \sin \theta ~e^{i\varphi}\\
i \sin \theta ~e^{-i\varphi} & \cos \theta\\
\end{pmatrix}.
\label{SM}
\ee
with time-independent elements.
For such impurity, without a loss of generality, one can put $\varphi=0.$   The effects  caused by slow  random evolution of
  $\mathbf M$
  in the external magnetic bath will be briefly discussed in Sec.~\ref{dephasing}.

The  resonant behavior of the  conductance  appears already in the perturbative regime, when impurity is weak and $\theta$ is sufficiently small. Then, one can only
keep the corrections   to the conductance up  to second order with respect to $\theta.$
Within this approximation one can consider only the ballistic processes and  processes with one and two backscattering events. Then,  the transmission coefficients for spin  $|\!\uparrow\rangle$  and $|\!\downarrow\rangle$ are  given by
 \be
\mathcal  T_\alpha= \left \langle \left |A^{(a)}_{\alpha} +A^{(b)}_{\alpha} \right |^2+ \left |A^{(c)}_\alpha \right |^2 \right \rangle_\varepsilon,   \qquad  \alpha = ~\uparrow,\downarrow
 \label{Tup}
 \ee
 Here $\langle \ldots \rangle_\varepsilon$  stands for  thermal averaging. Index  (a) corresponds to processes without backscattering (see,  Fig.~\ref{proc1}a and Fig.~\ref{proc2}a  for such processes with spin $\uparrow$ and $\downarrow,$ respectively).  Index
   (b)  corresponds   to the  paths with two backscattering events (see,  Fig.~\ref{proc1}b  for  $\alpha= \uparrow$ and Fig.~\ref{proc2}b  for  $\alpha=\downarrow$). Finally, contributions marked by  (c)  describes  spin-flip processes with a single backscattering act (Fig.~\ref{proc1}c and Fig.~\ref{proc2}c for different spin polarizations). As seen from Eq.~\eqref{Tup},  the      processes in which the outgoing spin polarization is parallel to the incoming  one [processes (a) and (b)]  interfere,  while spin-flip processes (c)   decouple.   Notice that both   $\mathcal T_\uparrow $ and $ \mathcal T_\downarrow$ depend on position of impurity. Hence,   $\mathcal T_\uparrow \neq \mathcal T_\downarrow.$

   One can calculate transmission coefficient by direct summation of amplitudes of processes shown in Figs.~\ref{proc1} and \ref{proc2} with the subsequent thermal averaging.
    This calculation is straightforward and is similar to   calculation of tunneling conductance of AB interferometer made of conventional materials under  the condition \eqref{TggD} \cite{dmitriev}, \cite{Shmakov2013}.  We thus  relegate it to Appendix
    \ref{formulas}.  The result reads
%

\be
\mathcal T=\frac{2\gamma}{1+\gamma^2 } -\frac{16 \gamma^3 A_\gamma ~ \theta^2}{1-\cos(4\pi\phi) +32 \gamma^2 B_\gamma}.
  \label{T-pert}
 \ee
 Here we expressed  tunneling amplitudes $r$  and $t$ in terms of $\gamma$ [see Eq.~\eqref{rt}] and introduced coefficients
\be
A_\gamma= \frac{1+6\gamma^2 +\gamma^4}{(1+\gamma^2)(1-\gamma^2)^4},~B_\gamma=\frac{(1+\gamma^2)^2}{(1-\gamma^2)^4},  \label{Agamma}\ee
which both  depend solely on the tunneling transparency.
 Equation \eqref{T-pert} gives the transmission coefficient for arbitrary tunneling coupling to the leads and weak coupling to magnetic impurity. The latter is taken into    account   perturbatively. We see that impurity induces   dependence of $\mathcal T$ on the  flux.

 Let us consider the limiting cases of Eq.\ \eqref{T-pert}. The  strong tunneling coupling (almost open ring with $r\to 1$, $t \to 0$) corresponds to    $\gamma \to 1$, see Eq.~\eqref{rt}.
From Eq.~\eqref{T-pert} we see that the dependence on magnetic flux  in this case is  very weak:
\be
\mathcal T\approx 1-\delta \mathcal T - \theta^2 t^4\cos(4\pi\phi),\qquad \text{for} \quad  t \to 0.
\ee
where
$\delta \mathcal T \approx  \theta^2/2 +2t^2$
is flux-independent correction.
Hence, for strong tunneling coupling the magnetic impurity leads to a small correction smoothly dependent on $\phi.$  This dependence is similar to  harmonic dependence on the AB phase predicted in Ref.~\cite{Dolcini2011}  for $T=0$  and slightly different geometry of the setup.

By   contrast,  in the opposite case of almost closed ring weakly coupled to the leads,  i.e. $\gamma \to 0,$   the  dependence   on the flux is very sharp:
\be
\mathcal T\approx 2\gamma \left[ 1 - \frac{8\gamma^2 \theta^2}{\displaystyle 1-\cos(4\pi\phi) + 32 \gamma^2 }\right], \quad \text{for} \quad  \gamma \to 0.
\label{T-pert1}
\ee
We see  that    $\mathcal T$
shows
narrow antiresonances as a function of $\phi$ of width $ \sim \gamma$  at $\phi=n$ and $\phi=n+1/2. $
Equation \eqref{T-pert1} is valid for arbitrary $\phi$ provided  that $\theta \ll \gamma \ll1.$  In the  vicinity of the resonances,   $\delta \phi = \phi - n \ll 1 $ or $\delta \phi = \phi - (n+1/2) \ll 1$, one may obtain
 non-perturbative  in $\theta$ solution which is valid for  $\theta \ll1 $  and arbitrary relation between    $\gamma$ and $ \theta$ (see Appendix):
%
\be
\mathcal T\approx 2\gamma \left[ 1 -\frac{\theta^2}{4} \frac{\gamma^2}{\displaystyle \gamma^2+ \left({\pi\delta \phi}/{2}\right)^2 + \left({\theta}/{4}\right)^2}\right].
\label{T-non-pert}
\ee
Thus,
 the non-perturbative effects lead  to appearance of the   additional contribution  $\theta^2/16$
in the denominator of the
resonant term in \eqref{T-non-pert}.  Physically, this corresponds to the broadening  of  the antiresonances because of multiple coherent scattering events.

\begin{figure}[ht]
\centerline{\includegraphics[width=\linewidth]{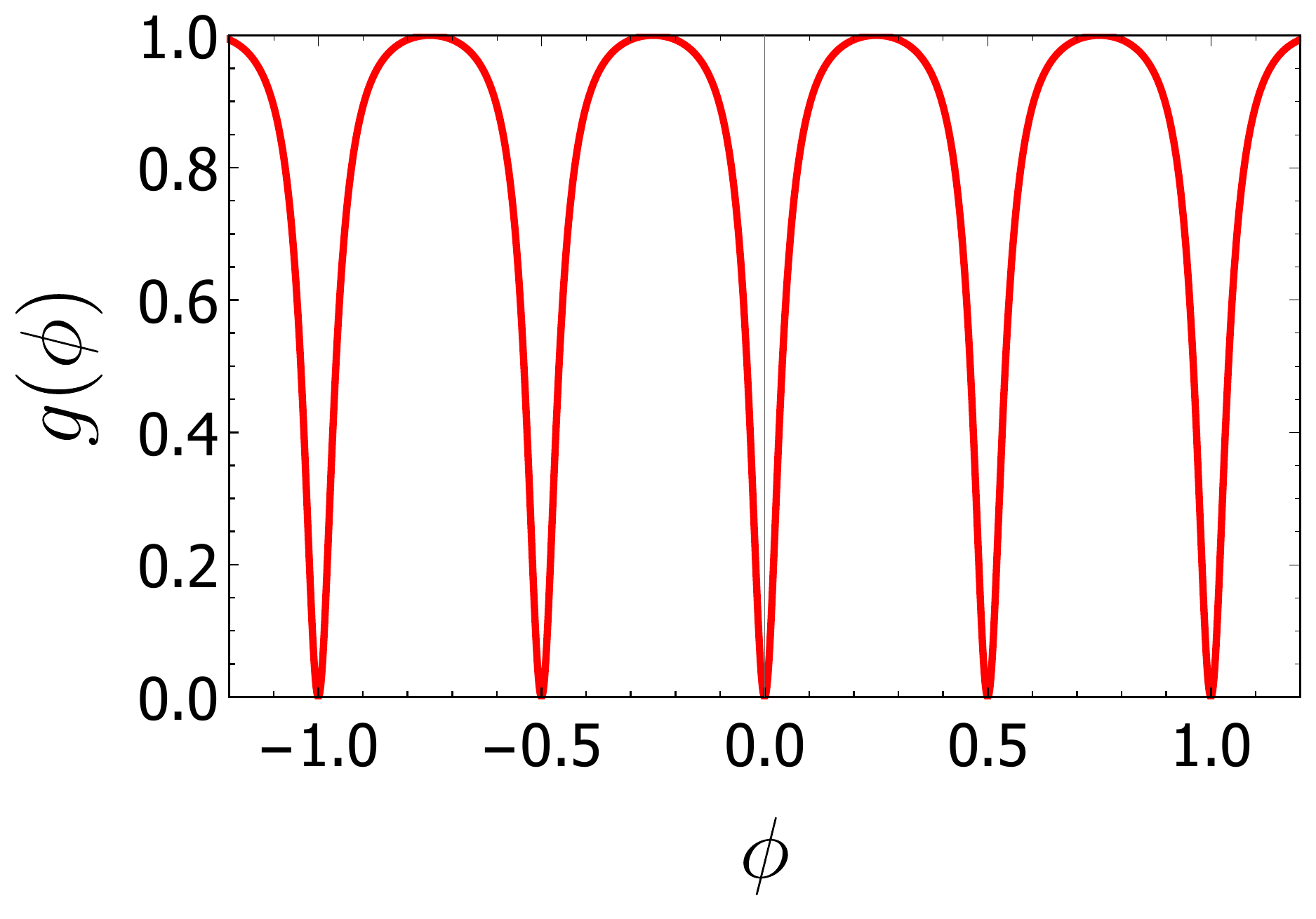}}
\caption{ The sharp antiresonances in the normalized conductance $g(\phi)$, Eq.\  \eqref{gPhi}.
  }
\vspace*{0.3cm}
\label{Fig:GvsPhi}
\end{figure}

We thus find  that the transmission coefficient and, consequently,  the conductance, $G=G(\phi)$, have minima at $\phi = n/2$, and maxima at $\phi = 1/4 + n/2$ with integer $n$.  Instead of  $G(\phi)$ it is
convenient to  introduce the following normalized function
\begin{equation}
g (\phi)  = \frac{ G(\phi)  -  G (0)} {G(1/4)  - G (0)}  \,.
\label{gPhi}
\end{equation}
which is plotted in Fig.\ \ref{Fig:GvsPhi}.

We see that  transmission coefficients is universal in a sense that it
does  not depend on details of the setup.  Such universality demands for a clear physical explanation.  Although the Berry's phase containing information about   structure of the helical edges   drops out  under condition \eqref{TggD}, as we explained above  in the end of Sec.~\ref{basic}, one also  needs  to  explain why  the conductance is  exactly periodic with the period $1/2$  and  does not depend on the  position of the magnetic impurity and relation between $L_1$ and $L_2.$ The next section is devoted to clarification of this issue.

\section{Physical interpretation: Coherent enhancement of the scattering off magnetic impurity} \label{phys}
In this Section,  we demonstrate that the obtained result can be derived  by taking into account the  coherent enhancement  of   backward scattering by magnetic impurity  at  integer and half-integer values of  $\phi$, which is accompanied by a decrease of backward scattering (or, equivalently, increase of the forward scattering) at other values of the flux.
The effect is fully analogous  to coherent enhancement of backscattering caused by weak localization (see Ref.\  \cite{classics} for review). As we will see, this analogy significantly helps to  explains universality of the result.

This analogy can be clarified by analyzing Figs.~ \ref{proc1}  and  \ref{proc2}.     Trajectories (a) and (b) with equal $n$ contain a loop (marked by red color) where  both trajectories   start  from  magnetic impurity and  make $n$ full revolutions in the opposite directions, clockwise for  the   trajectory (a) and   counterclockwise for the trajectory (b).    This is the typical weak localization loop.  The only essential difference as compared to weak localization  in   disordered  diffusive systems is that  in the case of helical edges the returns to impurity are ballistic.
  As illustrated  in Figs.~\ref{proc1}c  and \ref{proc2}c,  two spin-flip processes with  a given winding number but different  sequence  of backward and forward scattering events can  also contain a closed loop  (marked by red color) passed in the opposite directions.
   Two processes where  a closed loop is passed in the opposite directions      have equal lengths  and the same final spins,  so that  they interfere and  the corresponding flux-sensitive interference contribution  is not affected by  the thermal averaging.

Let us now demonstrate that the quantum correction to the conductance is {\it  fully} determined by the "weak localization" processes.  The simplest way to do this is to use the approach of
  Ref.~\cite{non-back},  where it was shown that the
 the effect of weak localization in 2D electron gas can be fully incorporated into the renormalization of the classical differential  cross-section  on a single impurity.
 Hence, one can describe weak localization effect by classical Drude-Boltzmann equation with the differential cross-section renormalized by the quantum correction.  The latter is   proportional to   the so-called  Cooperon, which is given by the sum of the maximally crossed diagrams.
 %

Next, we  demonstrate that a similar description is applicable to our problem.  To this end, we first consider  transport in our setup  within the classical approximation, which implies that   the magnetic impurity is described by the  forward  and backward scattering  probabilities:
\begin{equation}
t_{\rm f}^0=\cos^2\theta, \qquad  t_{\rm b}^0=\sin^2\theta.
\label{eq:21}
\end{equation}
 The classical transmission  coefficient can be presented as the sum over contributions from classical trajectories    propagating  clockwise and counterclockwise and experience collisions with probabilities \eqref{eq:21}.  In the perturbative regime,  the result  reads (see Appendix)
 %
%
%
\be
\mathcal T_{\rm cl} \approx \frac{2\gamma}{1+\gamma^2} - \frac{2\gamma^2 \theta^2}{(1+\gamma^2)^2}.
\label{Tcl-pert}
\ee
This equation does not contain contribution from interfering trajectories and, hence, is flux-independent. One can check that  it can be obtained from  Eq.~\eqref{T-pert} for quantum transmission coefficient  by averaging over the flux:
\be
\mathcal T_{\rm cl}= \langle  \mathcal T  \rangle_\phi.  \label{TclTav}
\ee

\begin{figure}[ht]
\centerline{\includegraphics[width=0.9\linewidth]{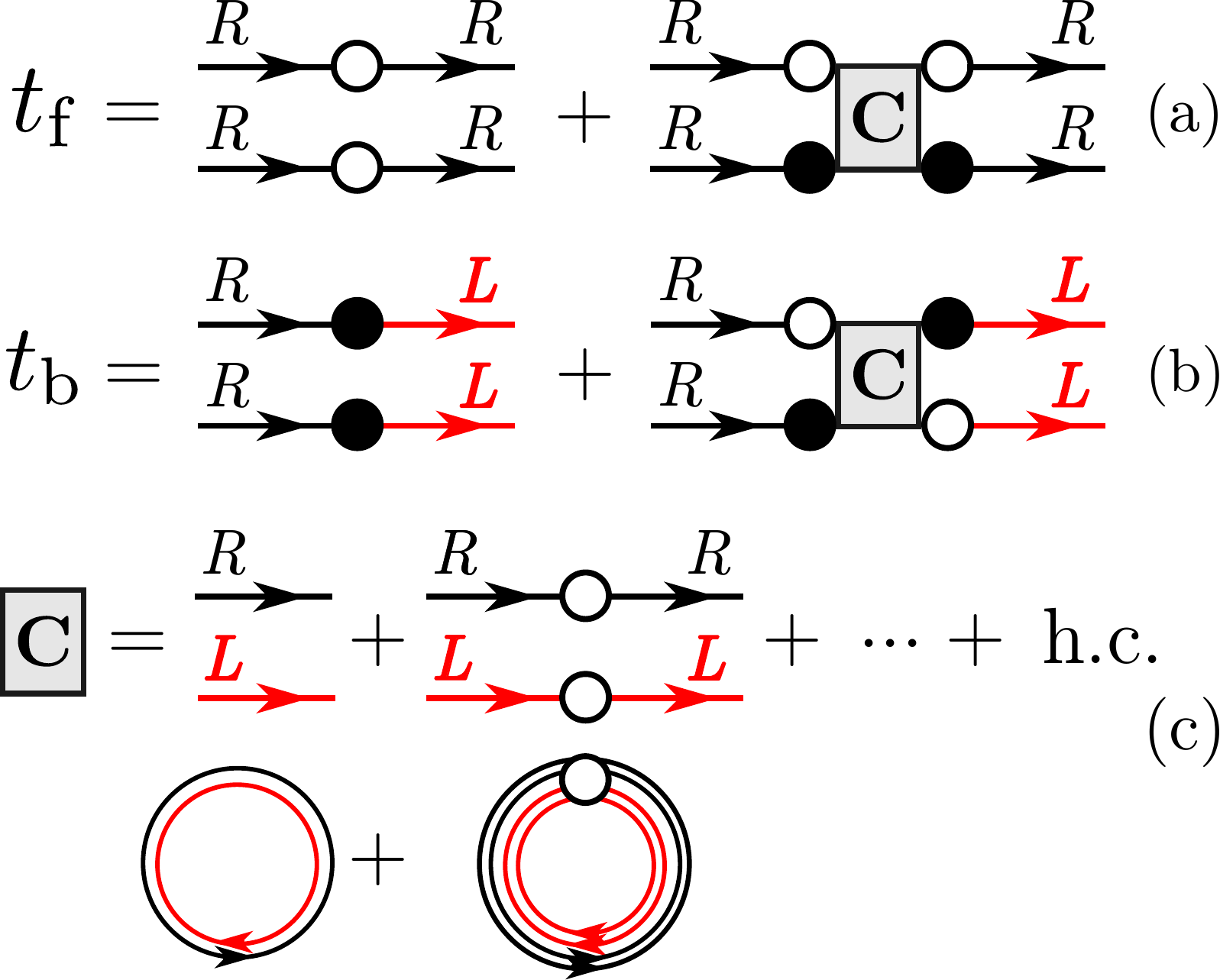}}
\caption{Forward (a) and backward (b) single scattering events   [left panels of (a) and (b), respectively]  and
coherent scattering processes  [right panels of (a) and (b)] leading to   enhancement  of scattering probabilities  due to ballistic returns to the magnetic impurity.   Probabilities of both coherent processes are proportional to the ballistic Cooperon (c)   which is given by the sum  over  closed ballistic loops formed by   trajectories propagating in the opposite directions  and
 having  different winding numbers.  Open circles and  black dots describe amplitudes of single  forward and backward scattering events, respectively.  The contributions of coherent processes  [right (a) and (b)  panels] coincide by absolute value, while the   phase factor $i$ arising in each backscattering act enters in different way,  $\delta t_{\rm f}=\cos^2\theta(-i\sin \theta)^2 C,$ $\delta t_{\rm b}=\cos^2\theta (i\sin \theta)(-i\sin \theta) C, $   which ensures  the probability conservation: $\delta t_{\rm f}+\delta t_{\rm b}=0.$          }
\label{Fig-a-b-c}
\end{figure}

As a next step, we  will show that the difference between quantum and classical results [Eq. ~\eqref{T-pert} and Eq.~\eqref{Tcl-pert}, respectively]   can be {\it fully} expressed in terms of
quantum weak localization corrections to the  classical
scattering  probabilities $t_{\rm f}^0$ and $t_{\rm b}^0.$   The latter  describe single scattering processes  by magnetic impurity shown in the left panels of Figs.~\ref{Fig-a-b-c} (a) and  (b),  respectively.
Let us  consider two processes involving multiple scattering by magnetic impurity  after some number of
ballistic returns.   First process is shown in Fig.~\ref{Fig-a-b-c}a (right panel)  and describes the  first order quantum correction to
 $t_{\rm f}^0.$   The first order   correction  to  the probability of the   backward scattering  is shown  in  the right panel of Fig.~\ref{Fig-a-b-c} (b).  Both processes  involve ballistic returns  to  the magnetic impurity  after  a number of  revolutions around the ring in which the  electron wave splits into two parts passing the setup in the opposite directions with equal winding  numbers.  Summing over winding numbers, we find ``ballistic Cooperon''
%
\be
{\cal C}=\frac{t^4 e^{4 i\pi\phi}}{1- t^4\cos^2\theta e^{4i\pi\phi} }+\frac{t^4 e^{-4i\pi\phi}}{1- t^4\cos^2\theta e^{-4i\pi\phi} },
\label{cooperon}
\ee
which  represents the contribution of the processes shown in Fig~\ref{Fig-a-b-c}(c).
Notice that
\begin{equation}
\langle  \mathcal C  \rangle_\phi =0 \,.
\label{eq:28a}
\end{equation}

%
%
One can easily calculate contributions from these processes to scattering probabilities
\be
\delta t_{\rm f }=-\cos^2\theta \sin^2\theta~ {\cal C},\quad   \delta t_{\rm b }=\cos^2\theta \sin^2\theta~{\cal C}.
\label{dtt}
\ee
As seen, these equations obey probability conservation law:
\be
\delta t_{\rm f }+\delta t_{\rm b }=0.
\ee
The total scattering probabilities incorporating first order  quantum corrections read
\be
t_{\rm f }= t_{\rm f }^0 + \delta t_{\rm f },\quad  t_{\rm b }= t_{\rm b }^0 + \delta t_{\rm b }.
\ee

\begin{figure}[ht]
\centerline{\includegraphics[width=\linewidth]{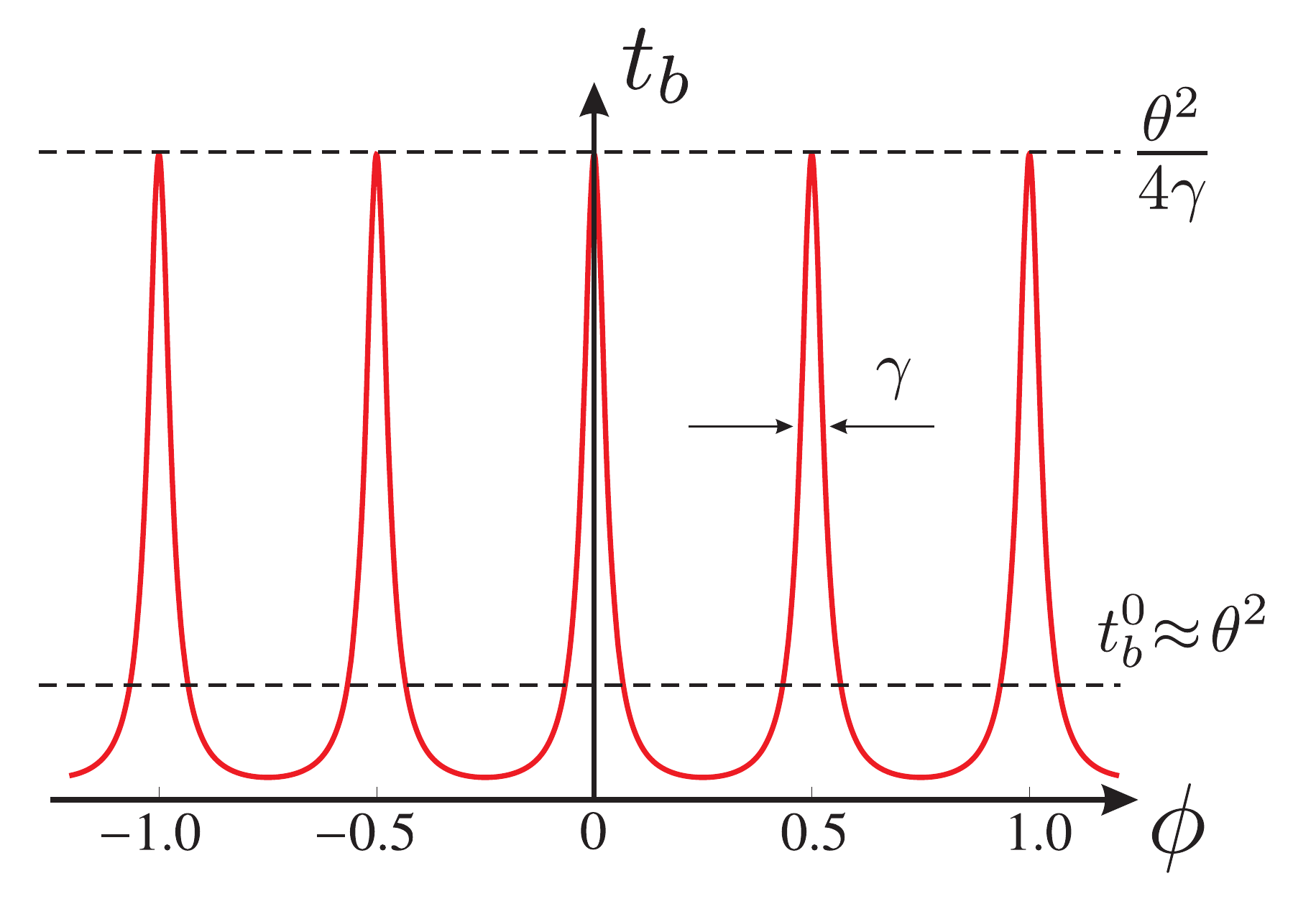}}
\caption{Dependence of total backscattering probability $t_{\rm b}$  on magnetic flux for weak tunneling coupling ($\gamma \ll 1$). This probability is strongly  enhanced in the vicinity of the resonances  and suppressed outside resonance region as compared to the classical flux-independent probability $t_{\rm b}^0$ shown by dashed line.  }
\vspace*{0.3cm}
\label{Fig-total}
\end{figure}

In the  limit of small $\theta,$  from Eqs.~\eqref{cooperon} and \eqref{dtt} we find
\be
 \delta t_{\rm b} \approx  \theta^2 {\cal C}|_{\theta=0}.
\label{dtb-theta}
\ee
Since ${t_{\rm b}^0} \simeq \theta^{2}$ at small $\theta$, see Eq.\ \eqref{eq:21},   the total factor appearing due to the coherent scattering is given by
\be
\frac{t_{\rm b} }{t_{\rm b}^0}\approx 1+{\cal C}|_{\theta=0}=\frac{8\gamma (1+\gamma^2)^2 A_\gamma}{32\gamma^2 B_\gamma + 1-\cos(4\pi\phi)},
\ee
where $A_\gamma$ is given by Eq.~\eqref{Agamma}.
For $\gamma \ll 1,$  this equation becomes
 \be
\frac{t_{\rm b}  }{t_{\rm b}^0}\approx \frac{8\gamma }{32\gamma^2  + 1-\cos(4\pi\phi)}.
\ee
Exactly at the resonances, the backscattering probability is enhanced by the  factor
\be
\left(\frac{t_{\rm b}  }{t_{\rm b}^0}\right)_{\rm max} =\frac{1}{4\gamma } \gg 1,
\ee
as compared to the classical backward probability.  On the other hand, the classical backscattering  probability is suppressed  beyond the resonance region.
  The dependence   $t_{\rm b} (\phi) $  is shown in Fig.~\ref{Fig-total}.

The dependence of  the quantum conductance $\mathcal T$ on $\phi$ at $\theta \to 0$ is obtained by replacing $\theta^{2}$ in  the classical expression, Eq. \eqref{Tcl-pert}, with its renormalized  value  $\theta^{2} \to \theta^{2} \left({t_{\rm b} }/{t_{\rm b}^0} \right) = \theta^{2} (1+{\cal C}|_{\theta=0}).$  Evidently, the relation \eqref{TclTav} is fulfilled due to   Eq.~\eqref{eq:28a}.

Concluding here, we  observe that the effective backward scattering is dramatically enhanced in vicinity of integer and half-integer values of the magnetic flux and is suppressed at other values of the flux.

 Based on derivation presented above, we can now explain the universal behavior of the  conductance.
 The universality is due to several physical  reasons:
 \begin{itemize}
   \item
   Multiple returns to magnetic impurity---the effect responsible for coherent enhancement of magnetic disorder---are only sensitive to the total length of the edge and independent of other details  of the interferometer, e.g., to relation between  $L_1$ and $L_2$ and position of the magnetic impurity.

   \item The exact periodicity of the conductance with the period $1/2$
   is a specific property of weak localization in a systems where the area  covered by  the   return loops    is fixed with the precision higher than the squared magnetic length \cite{aronov}.

   \item The Berry's phase, which encodes information about geometry of the setup   drops out from the high-temperature transmission coefficient.
    \end{itemize}
Actually, first two  statements in this list are less trivial than it seems at first glance. The key point here is the
        the absence of  backscattering by   non-magnetic contacts in helical edge.
         Due to this property the quantum correction to the conductance can be  {\it fully} expressed in terms of weak localization correction.  For non-helical  interferometers, where backscattering on contacts is inevitable, there are other interference contributions  which  are not  described in terms of weak localization correction and  lead to the violation of periodicity with the period $1/2$ \cite{comment}.

\section{Dephasing induced by dynamics of magnetic impurity}
\label{dephasing}
One of the key problems related to efficiency of any interferometer   is the dephasing caused by  inelastic  processes.  Here we briefly discuss one of the possible mechanisms of dephasing. A more detailed analysis will be presented elsewhere \cite{niyazov-to-be-published}.

The dominating  mechanism of dephasing in low dimensional systems is usually connected with the electron-electron interaction. For tunneling AB interferometers made of conventional    (non-helical) single-channel  ballistic wires, the  effect of  interaction  for $T\gg \Delta$  was studied  in Ref.~\cite{dmitriev} (see also Refs.~\cite{Dmitriev2015}, \cite{SCS2017}).    It was found  that    antiresonances in  the dependence  $G(\phi)$   split into  a series of narrow peaks
separated by distance $ \alpha$ with a smooth envelope   of the width $\alpha \sqrt {T/\Delta}$, where $\alpha$ is the dimensionless interaction constant. Each peak is broadened by dephasing within the  width $\gamma T/\Delta,$ which turns out to be interaction-independent  and gets suppressed with closing the interferometer ($\gamma \to 0$). The underlying physics is related to  interaction of the tunneling electrons with so-called zero-mode fluctuations, i.e.  tunneling-induced fluctuations of total numbers  of right- and left-moving electrons, $N_R$  and  $N_L$, respectively.

Similar  dephasing would also occur in the case under discussion \cite{NRNL}.  There is however, a  competing contribution   to dephasing caused by chaotic dynamics of impurity moment $\mathbf M. $ The latter mechanism  can  be more effective in an  almost closed interferometer of sufficiently large size $L.$   The random behavior  of  $\mathbf M (t) $    can occur for a number of reasons. First of all, the  dynamics of  $\mathbf M (t) $ appears due to  the exchange interaction with the  thermal bath of electrons  in the helical edge modes. This mechanism was recently discussed in Refs.~\cite{kur3} for the case of  infinite edge, where a general equation for relaxation of averaged moment $\langle \mathbf M (t)\rangle $ was derived.  Simple estimates show that for a ring-shaped edge of finite length $L$ weakly coupled to thermalized  leads the theory developed in  \cite{kur3} should be essentially modified. Most importantly, zero mode fluctuations come into play and dramatically change the dephasing action in close analogy with the effect of the electron-electron interaction \cite{dmitriev}. This would lead to a  non-trivial  dynamics of $\mathbf M (t) .$

Here, we confine ourselves to the analysis of   much simpler mechanism, namely, we assume that  a slow random dynamics of     $\mathbf M (t) $ is caused by the interaction with external magnetic bath, in particular, with thermal bath of bulk magnetic impurities. We also assume that this bath is isotropic, so that the correlation function describing the magnetic moment relaxation is given by
\be
\left\langle M_\alpha (0) M_\beta(t) \right \rangle=\frac{M^2}{3} \delta_{\alpha\beta} e^{-\Gamma t},
\label{MM}
\ee
where $\Gamma$ is the isotropic relaxation rate.

 The dephasing rate caused by relaxation \eqref{MM} can be calculated as follows.   We assume that the electron spin at the position of the magnetic impurity is directed along the $\hat z$ axis, and the components  of the classical vector $\mathbf M$  are given by:
$M_z=M \cos\chi , M_x=M \sin \chi \cos\varphi, M_y=M \sin \chi \sin\varphi. $ A slow evolution of $\mathbf M(t)$ is encoded in the time dependence of  angles $\chi=\chi(t)$ and $\varphi=\varphi(t).$ For fixed values of  $\chi$ and $\varphi$ the scattering matrix $\hat S_M$  is given by Eq.~\eqref{SM} with $\theta=g M \sin\chi,$ where $g \ll1$ is the exchange coupling constant.   Let us discuss the simplest situation, when the scattering on magnetic impurity is considered in the lowest order with respect to $\theta$, whereas $\chi$ is arbitrary.   The  correction to the backscattering for static impurity is given in this case by Eq.~\eqref{dtb-theta}.   In order to take into account the slow dynamics  of $\mathbf M$ this equation should be modified as follows:
\be
\delta t_{\rm b}= 2 {\rm Re}\sum \limits_{n=0}^{\infty}   \left(t^4 e^{4i\pi  \phi}\right)^{n+1} \left\langle \theta(0) \theta(t_n) e^{i(\varphi(0) -\varphi(t_n))}\right \rangle,
\ee
 where $t_n=2\pi n/\Delta.$  Next, we notice that    $\left\langle \theta(0) \theta(t_n) e^{i(\varphi(0) -\varphi(t_n)}\right \rangle=g^{2} \langle M_+(0)M_-(t)\rangle $ (with  $M_\pm=M_x \pm iM_y$) and use Eq.~\eqref{MM}. For $\Gamma \ll \Delta$ and $\gamma \ll 1,$ we find that Eq.~\eqref{T-pert} becomes
\be
\mathcal T \approx {2\gamma} -\frac{16 \gamma^3    \langle \theta^2 \rangle}{1-\cos(4\pi\phi) +32 (\gamma+\gamma_\varphi)^2 },
  \label{T-pert-final}
 \ee
where
\be
\gamma_\varphi =\frac{\pi \Gamma} {4\Delta}
\label{dephasing1}
\ee
is the dimensionless dephasing rate and $\langle \theta^2\rangle = g^2M^2 \langle\sin^2\chi \rangle=2g^2 M^2/3.$ The dephasing given by Eq.~\eqref{dephasing1} should dominate over interaction-induced one provided that relaxation rate $\Gamma$ is large as compared to interaction-induced splitting of antiresonances: $\Gamma \gg \alpha \sqrt{T\Delta}.$

\section{Conclusions}
We studied the transport
through the    tunneling Aharonov-Bohm   interferometer  formed by helical edge states
in the presence  of
a  magnetic impurity.
We  demonstrated that   at relatively high  temperatures,  $T \gg \Delta$ (which, for typical values of parameters, corresponds to temperatures higher than several kelvins),  the  tunneling  conductance of the interferometer  shows sharp antiresonances
The antiresonances are separated by distance $\varDelta \phi=1/2,$ which means that the tunneling conductance is periodic function of the magnetic flux with the period given by half of the flux quantum.
They are universal in a sense that their shape and position are not sensitive to geometry of the setup and to the  position of the magnetic impurity.
   We interpret  this  universal resonant behavior   as   a coherent enhancement of  backward scattering off magnetic impurity at integer and half-integer values  of flux, which is accompanied by suppression of the effective  scattering at other values of flux.  Both enhancement and suppression are    due to the interference of processes  involving   multiple returns (clockwise and counterclockwise) to magnetic impurity  after  revolutions  around interferometer  circumference --- the   phenomenon    similar to    weak-localization-induced enhancement  of backscattering.

   A very sharp dependence of the conductance  is very promising in view of possible   applications
  in the area of  extremely sensitive detectors of magnetic fields. Most importantly, the predicted dependence of the effective scattering probability on the magnetic flux  allows for flux-tunable control of the magnetic disorder.


\section{Acknowledgements}

The work of R.N. and V.K.  was supported by the joint grant of the Russian Science Foundation (Grant No. 16-42-01035) and the Deutsche Forschungsgemeinschaft (Grant No. MI 658-9/1).

\appendix*
\section{Technical details of calculations}
\label{formulas}

\subsubsection{Classical conductance}
The classical conductance of the interferometer with a magnetic impurity  can be  presented as a  following sum:
\be
\mathcal T_{\rm cl}=\frac{r^4}{2}\sum \limits_{n=0} \left[  J_1^{(n)} + J_2^{(n)}\right].
\ee
 Here $J_1^{(n)}$  ($J_2^{(n)}$)  is the  classical current, corresponding to particle which exit the interferometer  through the right contact  moving clockwise (counterclockwise)  after passing this contact $n$ times without scattering to the lead.
 Having in mind that the  impurity is  placed in the upper shoulder of the interferometer (see Fig.~\ref{fig:densityring}), one can easily find the following set of the recurrent equations for classical currents  $J_{1,2}^{(n)}$:
%
%
\be
\mathbf J^{(n+1)}=\begin{pmatrix}
t^4 \cos^2 \theta &  t^2 \sin^2 \theta \\
t^6 \sin^2 \theta  & t^4 \cos^2 \theta\\
\end{pmatrix}\mathbf J^{(n)}
\label{?}
\ee
with $\mathbf J^{(n)}=\left(J_1^{(n)},~J_2^{(n)}\right)$ and  $\mathbf J^{(0)} =(\cos^2\theta, 1+ t^2 \sin^2\theta).$

Solving these equations, expressing  $r$  and $t$  via $\gamma$ [see Eq.~\eqref{rt}],
we find
\be
\mathcal T_{\rm cl}=
2\gamma \frac{4\gamma + (\gamma^2-4\gamma+1) \sin^2\theta}{ 4\gamma (1+\gamma^2) + (1-\gamma)^4\sin^2\theta}.
\label{cl}
\ee
Expanding the result up to the second order with  respect to $\theta,$ we arrive at Eq.~\eqref{Tcl-pert} of the main text.

Next, we  calculate quantum conductance in  two different cases: the  very weak magnetic impurity  when calculations can be done perturbatively with respect to $\theta$ and the case when the magnetic impurity is relatively strong.

\subsubsection{Perturbative quantum analysis}
Here, we calculate sum of the amplitudes corresponding to  the processes shown in Figs.~\ref{proc1} and \ref{proc2}.   The energy dependence appears in the transmission amplitudes via
 the factor $\exp [i  (k L+\delta)].$  Due to the condition  $T\gg \Delta,$ this exponent rapidly oscillates   when
 energy changes within the temperature window around the Fermi energy.   To the exponential precision,
 one can thus replace the energy averaging by the averaging over the phase $\varphi=kL +\delta$ within the interval
 $0<\varphi<2\pi.$  Already on this stage the Berry's phase drops out.  Next one  can replace  $\exp (i\varphi)=z$ and reduce the
 averaging over $\varphi$ to the calculation of contour integral over $z.$ In all considered cases
 the  integrand in appearing integrals  has a simple pole structure and can be easily evaluated.

We start with deriving some auxiliary equations which allow one to perform thermal averaging.
 For any $a$ such that $|a|<1$ and any analytical function $F(z),$ we have
 \be
 \left \langle  \frac{F\left(e^{i\varphi}\right)}{1-a e^{-i\varphi}} \right \rangle_{\!\!\varphi} = F(a),
 \label{vspom1}
 \ee
  Using this equation, we find for any $b$ ($|b|<1$)   and $c$ ($|c|<1$):
 \begin{equation}
\begin{aligned}
  &  \left \langle \frac{e^{i n \varphi}}{\left(1-a e^{-i\varphi}\right)\left(1-b e^{i\varphi}\right)\left(1-c e^{i\varphi}\right) }
  \right \rangle_{\!\! \varphi}   \\
 &  = \frac{a^n}{(1-ba)(1-ca)} \,,   \\
\end{aligned}
\label{vspom2}
\end{equation}
and
 \begin{equation}
\begin{aligned}
 &   \left \langle  \left|\frac{1}{1-a e^{-i\varphi}}\right|^2 \left|\frac{1}{1-b e^{-i\varphi}}\right|^2
  \right \rangle_{\!\! \varphi}  \\
 &  = \frac{1-|a|^2|b|^2}{(1-|a|^2)(1-|b|^2)|1-b^*a|^2}  \,.  \\
\end{aligned}
\label{vspom3}
\end{equation}

  The summation of amplitudes of different interference processes can be obtained perturbatively in   $\theta$ provided that $\theta \ll {\rm max} (\gamma, 1).$
The tunneling amplitudes  of trajectories shown in Fig.~\ref{proc1}a,b  are given, respectively, by the following equations
  \begin{equation}
\begin{aligned}
 A^{(a)}_\uparrow &=   \frac{  r^2 \cos \theta e^{ i \varphi_+ L_1/L}}{1-t^2 \cos\theta e^{ i \varphi_+} } ,   \\
 A^{(b)}_\uparrow & =  \frac{  r^2  t^2 (i \sin\theta)^2  e^{i (\varphi_- +\varphi_+ L_1/L)}}{[1\!-\!t^2 \cos\theta  e^{ i \varphi_+} ]^2 [1\!- \!t^2  \cos\theta e^{ i \varphi_-} ]},     \\
\end{aligned}
\label{a-up12}
\end{equation}
 which are obtained by direct summation of tunneling amplitudes with different winding numbers.  Here
 \be
 \varphi_\pm=kL \pm 2\pi \phi
 \ee
 (we incorporated the Berry's phase into the dynamical phase).  The factor  $[1\!-\!t^2 \cos\theta  e^{ i \varphi_+} ]^{-1}$ in $A^{(a)}_\uparrow$  appears due to summation of contributions of  clockwise rotating ballistic trajectories   over winding number $n$.  The    denominator in $A^{(b)}_\uparrow $ stems from contributions of clockwise processes with different winding numbers both  before  and after backscattering,  together with  contributions  of counterclockwise scattering processes between two backscattering acts.

 The amplitude of the process shown in Fig.~\ref{proc1}c reads
 \be
  A^{(c)}_\uparrow =  \frac{  r^2  t ( i \sin\theta)  e^{i [\varphi_+s/L ~+~\varphi_-(L+s-L_1)/L]}}{[1- t^2 \cos\theta  e^{ i \varphi_+} ] [1- t^2  \cos\theta e^{ i \varphi_-} ]},  \label{a-up-3}
 \ee
 where $s$ is the distance from the left contact to the impurity position.

 The processes with opposite incoming spin are shown in   Fig.~\ref{proc2}.
The analytical expressions for corresponding  amplitudes  read
 \begin{equation}
\begin{aligned}
  A^{(a)}_\downarrow & =   \frac{  r^2 e^{ i \varphi_- (L-L_1)/L}}{1-t^2 \cos\theta e^{i \varphi_-} } , \\
 A^{(b)}_\downarrow & =  \frac{  r^2  t^4 (i \sin\theta)^2  e^{i (\varphi_+ +\varphi_- (2L-L_1)/L)}}{[1\!-\!t^2 \cos\theta  e^{ i \varphi_-} ]^2 [1\!- \!t^2  \cos\theta e^{ i \varphi_+} ]},    \\
 A^{(c)}_\downarrow & =  \frac{  r^2  t ( i \sin\theta)  e^{i [\varphi_+(L_1- s)/L +\varphi_-(L-s)/L]}}{[1- t^2 \cos\theta  e^{ i \varphi_+} ] [1- t^2  \cos\theta e^{ i \varphi_-} ]},
\end{aligned}
\label{a-down}
\end{equation}
 Substituting   Eqs.~\eqref{a-up12},  \eqref{a-up-3} ,  and  \eqref{a-down}
 into Eqs.~\eqref{Tup}, performing energy averaging  with the use of  Eqs.~\eqref{vspom1},  \eqref{vspom2}, and \eqref{vspom3} ,   expanding results over $\theta$ up to the second order  and using Eq.~\eqref{T},
   we  find after some algebra Eq.~\eqref{T-pert} of the main text.

\subsubsection{Non-perturbative quantum analysis} \label{app-nonpert}
For weak tunneling coupling, $\gamma \ll 1,$ one can derive non-perturbative in $\theta$ equation for transmission coefficient, which is valid for arbitrary relation between $\theta$ and $\gamma$  (but it is still assumed that $\theta \ll1$). To this end,
  one can use an approach developed in ~\cite{Shmakov2012,Shmakov2013} for conventional (non-helical) interferometers. For simplicity, we only consider the case of the interferometer with equal shoulders: $L_1=L_2.$

The electron tunneling  transmission amplitude   is a  matrix $\hat t= \hat t (\epsilon)$  in a spin space with elements $t_{\alpha \beta}=\langle \alpha | t | \beta \rangle .$  The transmission coefficient is expressed in terms of this amplitude    as follows:
\begin{equation}
{\cal T}( \epsilon) = \frac 12 \mbox{Tr } \hat t \hat t^\dagger.
\end{equation}
The tunneling trajectory with  $n$  full revolutions around the setup  has  a length $L_n  = L(n + 1/2).$ The transmission amplitude can be expressed as a  sum over $n$:
\begin{equation}
\hat t ( \epsilon) = \sum_{n=0}^\infty \hat \beta_n  e^{i k L_n}.
\end{equation}
The quantity $\hat \beta_n$ consists of two contributions: trajectories ending at the bottom semicircle, $\hat \beta^+_n$ and at the top semicircle, $\hat \beta^-_n$. The vector  constructed  out of  these two contributions,
\begin{equation}
\vec{ \hat{ \beta}}_n =
\begin{pmatrix}
\hat \beta^+_n \\
\hat \beta^-_n
\end{pmatrix},
\end{equation}
obeys a recurrence relation
 $$\vec{ \hat{ \beta}}_{n+1}= \hat A \vec{ \hat{ \beta}}_{n},$$
 where $\hat A$ is certain $n-$independent matrix. Then,   $\vec{ \hat{ \beta}}_{n}$ is given by
  $\hat \beta_n =  \vec{ \hat{ e}}\hat A^n \vec{ \hat{ \beta}}_0$, where
\begin{equation}
\vec{ \hat{ e}} =
\begin{pmatrix}
\mathbf{1} \\
\mathbf{1}
\end{pmatrix}.
\end{equation}
The transmission coefficient,
\begin{equation}
{\cal T}( \epsilon) = \frac 12 \text{Tr} \sum_{n,m=0}^\infty \hat \beta_n \hat \beta^\dagger_m e^{ik(L_n-L_m)},
\end{equation}
after energy averaging under condition $T \gg \Delta$ has only nonzero terms with $n=m$ and  is  expressed in terms of matrix $\hat A$ as follows
\begin{equation}
{\cal T} = \frac 12 \text{Tr} \sum_{n=0}^\infty |\hat \beta_n |^2=\frac 12 \text{Tr} \sum_{n=0}^\infty |\vec{ \hat{ e}}\hat A^n \vec{ \hat{ \beta}}_0|^2.
\end{equation}
Calculation of the  sum (see technical details in Refs.~\cite{Shmakov2012,Shmakov2013})   yields
a  matrix $ \hat B =(\mathbf 1 - \hat A \otimes \hat A^\dagger)^{-1}$ acting in the space of doubled dimension.  Hence, the problem is reduced to a straightforward (albeit cumbersome) calculation of matrix elements of $\hat B.$

Because of the helical nature of the edge state the  matrix $\hat A$ is simplified since    we  assume that counterclockwise electron propagation mode with the spin up as well as clockwise mode with spin down are absent:
\begin{equation}
 \hat A =
\begin{pmatrix}
 0 & 0 & 0 & 0 \\
 0 & -e^{2 i \pi  \phi } t^2 \cos \theta  & -e^{2 i k s} t^3 \sin \theta  & 0 \\
 0 & e^{-2 i k s} t \sin \theta  & -e^{-2 i \pi  \phi } t^2 \cos \theta & 0 \\
 0 & 0 & 0 & 0 \\
\end{pmatrix}.
\end{equation}
The initial amplitudes read
\begin{equation}
\begin{aligned}
\hat \beta^+_0&=
\begin{pmatrix}
 0 & 0 \\
 -e^{i \pi  \phi } (1-t^2) & -e^{2 i k s+i \pi  \phi } t (1-t^2) \sin \theta \\
 \end{pmatrix}, \\
\hat \beta^-_0&=
 \begin{pmatrix}
 0 & -e^{-i \pi  \phi } (1-t^2) \cos \theta  \\
 0 & 0 \\
\end{pmatrix}.
\end{aligned}
\end{equation}
Using these formula, one can represent the transmission coefficient  as a ratio of two rather cumbersome functions of     $\theta,\gamma$ and $\delta \phi. $
In the limit $\theta,\gamma,\delta \phi \ll 1,$   one can expand both functions over   $\theta,\gamma$ and $\delta \phi $ up to the terms of  second order.  Doing so,  we arrive at  Eq.~\eqref{T-non-pert} of the main text. One can show that this equation is also valid for $L_1 \neq L_2.$


\begin{thebibliography}{80}
\bibitem{NRNL} An important modification of results of Ref.~\cite{dmitriev} in the case of helical ring coupled to magnetic impurity is related to the
fact that  $N_R$ and  $N_L$ are non conserved even in the fully closed setup ($\gamma=0$) because of the impurity-induced backscattering.
\end{thebibliography}

\end{document}